\documentclass{pasj02}

\Received{17-Nov-2024}
\Accepted{13-Dec-2024}
\Published{$\langle$publication date$\rangle$}

\usepackage{url,xcolor,CJKutf8,acronym,ulem,braket}
\usepackage[switch,mathlines]{lineno}

\newcommand{\msun}{\mathrm{M}_\odot}

\acrodef{srga}[SRGA J1444]{SRGA J144459.2$-$60420}
\acrodef{lmxb}[LMXB]{low mass X-ray binary}

\begin{document} 

\title{ 
\LETTERLABEL 
Evidence of non-Solar elemental composition\\
in the clocked X-ray burster SRGA J144459.2$-$604207}


\author{Akira \textsc{Dohi}\altaffilmark{1,2}}
\email{akira.dohi@riken.jp}

\author{Nobuya \textsc{Nishimura}\altaffilmark{3,1,4}}
\author{Ryosuke \textsc{Hirai}\altaffilmark{1,5,6}}
\author{Tomoshi \textsc{Takeda}\altaffilmark{7,1}}
\author{Wataru \textsc{Iwakiri}\altaffilmark{8,1}}
\author{Toru \textsc{Tamagawa}\altaffilmark{1}}
\author{Amira \textsc{Aoyama}\altaffilmark{7,1}}
\author{Teruaki \textsc{Enoto}\altaffilmark{9,1}}
\author{Satoko \textsc{Iwata}\altaffilmark{7,1}}
\author{Yo \textsc{Kato}\altaffilmark{1}}
\author{Takao \textsc{Kitaguchi}\altaffilmark{1}}
\author{Tatehiro \textsc{Mihara}\altaffilmark{1}}
\author{Naoyuki \textsc{Ota}\altaffilmark{7,1}}
\author{Takuya \textsc{Takahashi}\altaffilmark{7,1}}
\author{Sota \textsc{Watanabe}\altaffilmark{7,1}}
\author{Kaede \textsc{Yamasaki}\altaffilmark{7,1}}


\altaffiltext{1}{RIKEN Cluster for Pioneering Research (CPR), RIKEN, Wako, Saitama 351-0198, Japan}
\altaffiltext{2}{Interdisciplinary Theoretical and Mathematical Sciences Program (iTHEMS), RIKEN, Wako, Saitama 351-0198, Japan}
\altaffiltext{3}{Center for Nuclear Study (CNS), The University of Tokyo, Bunkyo-ku, Tokyo 113-0033, Japan}
\altaffiltext{4}{National Astronomical Observatory of Japan (NAOJ), Osawa, Mitaka 181-8588, Japan}
\altaffiltext{5}{School of Physics and Astronomy, Monash University, Clayton, VIC 3800, Australia}
\altaffiltext{6}{OzGrav: The ARC Centre of Excellence for Gravitational Wave Discovery, Clayton, VIC3800, Australia}
\altaffiltext{7}{Department of Physics, Tokyo University of Science, Kagurazaka, Shinjuku-ku, Tokyo 162-8601, Japan}
\altaffiltext{8}{International Center for Hadron Astrophysics, Chiba University, Chiba 263-8522, Japan}
\altaffiltext{9}{Department of Physics, Kyoto University, Sakyo, Kyoto 606-8502, Japan}



\maketitle

\begin{abstract}
In February and March 2024, a series of many Type I X-ray bursts from the accreting neutron star SRGA J144459.2$-$604207, which has been identified by multiple X-ray satellites, with the first reports coming from INTEGRAL and NinjaSat. These observations reveal that after exhibiting very regular behavior as a ``clocked'' burster, the peak luminosity of the SRGA J144459.2$-$604207 X-ray bursts shows a gradual decline. The observed light curves exhibit a short plateau feature, potentially with a double peak, followed by a rapid decay in the tail-features unlike those seen in previously observed clocked bursters. In this study, we calculate a series of multizone X-ray burst models with various compositions of accreted matter, specifically varying the mass fractions of hydrogen ($X$), helium ($Y$), and heavier CNO elements or metallicity ($Z_{\rm CNO}$). We demonstrate that a model with higher $Z_{\rm CNO}$ and/or lower $X/Y$ compared to the solar values can reproduce the observed behavior of SRGA J144459.2$-$604207. Therefore, we propose that this new X-ray burster is likely the first clocked burster with non-solar elemental compositions. Moreover, based on the X-ray burst light curve morphology in the decline phase observed by NinjaSat, a He-enhanced model with $X/Y \approx 1.5$ seems preferred over high-metallicity cases. We also give a brief discussion on the implications for the neutron star mass, binary star evolution, inclination angle, and the potential for a high-metallicity scenario, the last of which is closely related to the properties of the hot CNO cycle.
\end{abstract}


\clearpage 

\section{Introduction}

On 26 February 2024, new observations of Type I X-ray bursts (hereafter, X-ray burst: XRB) from \ac{srga}, located in the Galactic disc, were conducted by several X-ray astronomical satellites, e.g., NICER \citep{2024ATel16474....1N,2024ApJ...968L...7N}, INTEGRAL \citep{2024ATel16485....1S}, MAXI \citep{2024ATel16483....1N}, IXPE \citep{2024arXiv240800608P}, Insight-HXMT \citep{2024ATel16548....1L}, and NinjaSat \citep{2024ATel16495....1T}. The observed bursts from \ac{srga} are considered to be a {\it clocked} burster, which shows a burst-to-burst uniform light curve profile and constant recurrence time. Thus, \ac{srga} is the sixth confirmed clocked XRB source after GS 1826$-$24 \citep{1999ApJ...514L..27U, 2001AdSpR..28..375C, 2003A&A...405.1033C, 2004ApJ...601..466G}, GS 0836$-$429 \citep{2016A&A...586A.142A}, IGR J17480$-$2446~\citep{2011MNRAS.418..490C}, MAXI J1816$-$195 \citep{2022ApJ...935L..32B,2024A&A...689A..47W} and 1RXS J180408.9$-$342058 \citep{2017MNRAS.472..559W,2019ApJ...887...30F}.

Type I XRBs are thermonuclear explosions triggered by nuclear ignition of hydrogen and helium in the accreted layer on the neutron star (NS) surface. Observations of XRBs provide various information relevant to nuclear astrophysics and the physical properties of low-mass X-ray binaries. Among more than 100 XRB sources observed so far \citep{2020ApJS..249...32G}, the most suitable sites for constraining the physics of low-mass X-ray binaries are the clocked bursters due to their uniform features (for a review, see \cite{2017PASA...34...19G} and references therein). Since the first modeling of the well-known clocked burster GS 1826$-$24 \citep{2007ApJ...671L.141H}, observations of clocked burster have been used to constrain physical properties in accreting NSs, companion stars and surface nuclear burning (e.g. \cite{2016ApJ...819...46L,2018ApJ...860..147M,2020PTEP.2020c3E02D, 2020MNRAS.494.4576J, 2021PhRvL.127q2701H, 2021ApJ...923...64D, 2022ApJ...937..124D, 2022ApJ...929...72L, 2022ApJ...929...73L}). 
Most previous studies on clocked bursters focused on the hydrogen-rich compositions (but see \cite{2019ApJ...872...84M,2020MNRAS.494.4576J}), which actually accounts for the long burst duration of GS 1826--24. Note that clocked bursters observed so far showed that the luminosity \textit{slowly} decays around a minute after it linearly increases.

\begin{figure}
 \centering
 \includegraphics[width=0.75\linewidth]{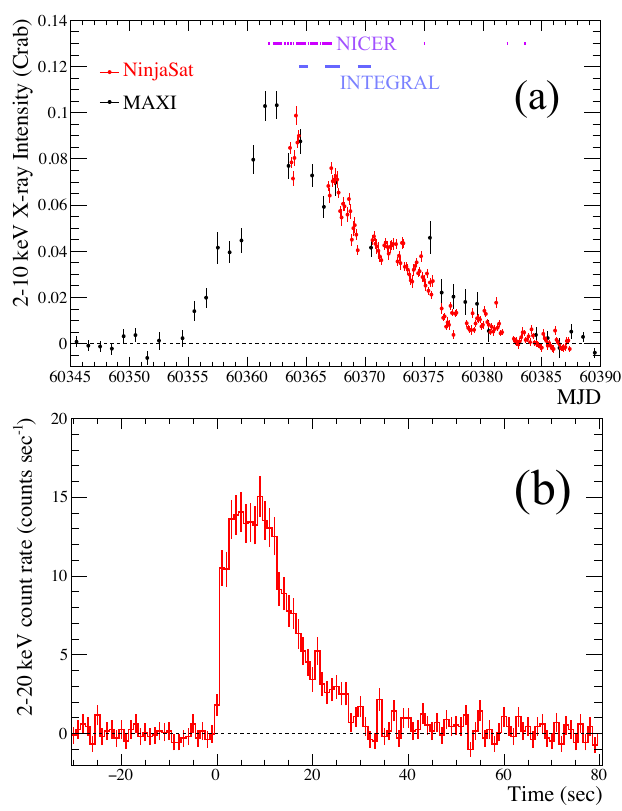}
\caption{2--10 keV light curves of SRGA J1444, monitored by NinjaSat (red) and MAXI (black), with bin sizes of 3~h and 24~h, respectively. Horizontal bars indicate the time intervals covered by observations of NICER (purple) and INTEGRAL (blue).
(b) The averaged burst profile in 11 profiles observed by NinjaSat~\citep{2025PASJL}. 
}
 \label{fig:obs}
\end{figure}

The main observed features of \ac{srga} are shown in figure~\ref{fig:obs}. The MAXI light curve (black) indicates a rise and fall in the persistent X-ray flux, indicating variability in the mass accretion rate over a duration of about one month (see Section~\ref{sec:parameters}). XRBs were observed during and after the peak, as indicated by the colored lines in figure~\ref{fig:obs}. The bursts were periodic around the peak of the persistent flux, suggesting that \ac{srga} is a clocked burster, with periodic recurrence times of $1.69$~h \citep{2024ATel16485....1S}. In the later phase, the luminosity of the \ac{srga} bursts gradually decline, and the recurrence time increases, as observed by IXPE \citep{2024arXiv240800608P} and NinjaSat \citep{2025PASJL}. \citet{2025PASJL} have reported successful observations of 11 burst events for around two weeks in the decline phase, with the last two occurring consecutively, separated by $\Delta t = 7.909~{\rm h}$. Unlike other typical clocked bursters, such as GS 1826$-$24 \citep{2017PASA...34...19G}, the bursts in \ac{srga} exhibit a plateau peak, possibly featuring a double peak, followed by a rapid decay with an exponential decay timescale of $\tau_e \sim 9~{\rm s}$ (figure~\ref{fig:obs}b and see also \cite{2024A&A...690A.353M} and \cite{2024ApJ...968L...7N}).

In the present study, we theoretically investigate \ac{srga} to constrain the physical properties of the low-mass X-ray binary system. We calculate a series of multizone XRB models using a 1D spherically symmetric general relativistic stellar evolution code, \texttt{HERES} \citep{2020PTEP.2020c3E02D}. With representative mass accretion rates, we compare our models with a variety of elemental {\it compositions} of the accreted matter, i.e., hydrogen ($X$), helium ($Y$), and heavier CNO elements or metallicity ($Z_{\rm CNO}$). We find that the model with higher $Z_{\rm CNO}$ or lower $X/Y$ compared to the solar compositions can reproduce the observations, including the recurrence time $\Delta t$ with appropriate choices of the accretion rate.

This paper is structured as follows: In Section~\ref{sec:method}, we summarize the numerical methods for XRB calculations, including the initial conditions and the adopted binary parameters. The results are presented in Section~\ref{sec:res}. Section~\ref{sec:summary} provides a summary and discussion.

\section{Methodology: XRB modeling}
\label{sec:method}

\subsection{XRB code: \texttt{HERES}}

In modeling theoretical light curves of \ac{srga}, we utilize a general relativistic stellar evolution code, \texttt{HERES} (Hydrostatic Evolution of RElativistic Stars) described in \citet{1984ApJ...278..813F,2020PTEP.2020c3E02D}, which simultaneously solves the Tolman-Oppenheimer-Volkoff equation along with thermal transport equations, accounting for nuclear burning and neutrino cooling. The validity of \texttt{HERES} has been thoroughly verified by comparing its results with \texttt{MESA} models \citep{2023ApJ...950..110Z}. SRGA J1444 does not exhibit photospheric radius expansion \citep{2024A&A...690A.353M}. This allows us to safely utilize \texttt{HERES} that assumes hydrostatic equilibrium condition\footnote{Although \cite{2024arXiv241205779F} has recently suggested the occurrence of photospheric radius expansion, our modeling by HERES is still valid in case of weak photospheric expansion~\citep{1983ApJ...267..315P} }.

The \texttt{HERES} code solves the thermal evolution of accreting NSs, considering realistic equations of state (EOSs) and heating and cooling sources. We adopt the Togashi EOS \citep{2017NuPhA.961...78T}, which has been constructed with realistic nuclear potentials at finite temperature, specifically for 1.4 $M_{\odot}$ NSs (with a radius of 11.6 km). Regarding NS cooling, we consider ``slow neutrino cooling processes'', mainly composed of the modified Urca process and Bremsstrahlung. Note that fast neutrino cooling processes, such as the Direct Urca process, are prohibited with the Togashi EOS \citep{2019PTEP.2019k3E01D}. For the heating in the NS crust, which consists of electron capture, neutron emission, and pycnonuclear fusion, we adopt the standard heating rates \citep{2008A&A...480..459H} to describe their energy generation.

Explosive nuclear-burning phases in the burst regimes are solved with a nuclear reaction network coupled with the hydrostatic evolution. Our approximated reaction network is only valid for $X/Y \gtrsim 1$, which satisfies our current parameter space. At the typical XRB density of $\rho \sim 10^{6}~{\rm gcm^{-3}}$, burning processes proceed from the triple-$\alpha$ reaction, through the hot CNO (bi-)cycle, and into the $\alpha p$ and $rp$ processes \citep[for a review]{2018JPhG...45i3001M}. To efficiently cover all nucleosynthesis processes, we implement an approximate reaction network with $88$ nuclei, ranging from n/p to ${}^{107}{\rm Te}$ (see \cite{2020PTEP.2020c3E02D} for details). All relevant reaction rates are sourced from JINA Reaclib. Note that our previous studies, such as \cite{2021ApJ...923...64D}, used JINA Reaclib version 2.0. Both versions are available in the JINA Reaclib Database\footnote{\url{https://reaclib.jinaweb.org}}.

\subsection{Accretion model parameters}
\label{sec:parameters}

Besides the physical properties of the NS, the input parameters for calculating XRBs are related to the values of the X-ray binary system, i.e., the accretion rate $\dot{M}_{-9}$ and the composition of the accreted matter: the mass fractions of hydrogen ($X$), helium ($Y$), and metallicity ($Z$). For the elemental composition, we consider the hydrogen-helium ratio ($X/Y$), and $Z$ is represented by the total CNO nuclei, $Z = Z_{\rm CNO}$.

In modeling the \ac{srga} bursts, we distinguish two phases. The first phase is the clocked burst phase, corresponding to the phase with high persistent flux ($\sim0.1~{\rm Crab}$). In this paper, we take the information of INTEGRAL of $\Delta t\simeq1.69~{\rm hr}$ and $\tau_e\simeq9~{\rm s}$, and the light curves by NICER, available under ObsID 6204190102 with MJD 60362.35007 and previously analyzed in \citet{2024ApJ...968L...7N}~\footnote{The light-curve profile remains similar even when using other NICER burst data.}. The NICER data were processed using NICERDAS version 11 as distributed with HEASOFT version 6.32. The second phase is where the persistent luminosity of \ac{srga} gradually declines, corresponding to when the NinjaSat and IXPE observations were taken. In the NinjaSat observations, the 10th and 11th XRBs occur consecutively, which gives $\Delta t_{\rm NinjaSat}=7.909~{\rm h}$~\citep{2025PASJL}. For the light curves, we take the averaged profile of the 10th and 11th XRBs, assuming a bin size of 1~s. The light curve by NinjaSat is calculated in the 2-20 keV energy band after subtracting persistent emission.

Generally, there is an empirical relation between the accretion rate and XRB recurrence time (see e.g., \cite{2016ApJ...819...46L}):
\begin{equation}
    \dot{M}_{-9}\propto\Delta t^{-\eta}
    \label{eq:eta}
\end{equation}
where the parameter $\eta$ is close to 1. In the case of \ac{srga}, $\eta$ is estimated to be $\sim0.8-0.9$~\citep{2024arXiv240800608P,2025PASJL}. From the observed $\Delta t$, one can constrain the accretion-rate ratio between the clocked phase and the later phase observed by NinjaSat (see figure~\ref{fig:obs}b) as roughly
\begin{equation}
    \frac{\dot{M}_{-9,{\rm NinjaSat}}}{\dot{M}_{-9,{\rm clocked}}}=\left(\frac{\Delta t_\mathrm{clocked}}{\Delta t_\mathrm{NinjaSat}}\right)^\eta
    \simeq 0.2 - 0.3 \,
    \label{eq:mdot}
\end{equation}
where $\dot{M}_{-9,{\rm clocked}}$ and $\dot{M}_{-9,{\rm NinjaSat}}$ are the mass accretion rates at the clocked phase and the later phase observed by NinjaSat, respectively. For the clocked burst phase, we adopt $\dot{M}_{-9,{\rm clocked}}= 3$ and $4$, which are appropriate $\dot{M}_{-9,{\rm clocked}}$ values for reproducing $\Delta t = 1.69~{\rm h}$ by \citet{2024ApJ...960...14D}~\footnote{Although \citet{2024ApJ...960...14D} considered the only case with $X/Y=2.9$, $\Delta t$ is not so changed by the $X/Y$ values considered here (see also Appendix~\ref{app}).}. Based on the constraint of equation~(\ref{eq:mdot}), we adopt $\dot{M}_{-9}=0.8$ and $0.9$ for the modeling of NinjaSat light curves.

In contrast to the accretion rate $\dot{M}_{-9}$, we adopt common values for the composition in both phases. In this study, we focus on the impact of composition by considering three cases: (i) the solar composition, with $X/Y  = 2.9 \equiv ({X/Y})_\odot$ and $Z_{\rm CNO} = 0.015 \equiv Z_\odot$~\citep{2019arXiv191200844L}, (ii) He-enhanced case, with $X/Y = 1.5 \simeq (X/Y)_\odot /2$ and $Z_{\rm CNO} = Z_\odot$ (iii) CNO-enhanced case, with $X/Y = (X/Y)_\odot$ and $Z_{\rm CNO} = 4~Z_\odot$, where the metallicity is composed of ${}^{14}{\rm O}$ and ${}^{15}{\rm O}$ in the ratio of mass fractions $7:13$ \citep{1983PASJ...35..491H}. Furthermore, we partially simulate additional metallicity cases, $Z_{\rm CNO}/Z_\odot = 2$ and $3.3$. 

To minimize the influence of compositional inertia \citep{2004ApJS..151...75W}, we prepare the XRB initial conditions by calculating the steady state with compressional heating \citep{2021ApJ...923...64D}. We discard the first 20 burst events for the early clocked phase and the first event for the later NinjaSat phase. For the surface boundary, we adopt the radiative-zero boundary condition at $M_r/M_{\rm NS}=10^{-16}$ for most cases \citep{1983ApJ...267..315P} \footnote{In the clocked burster phase, we confirm that the light curves are unchanged between $M_r/M_{\rm NS}=10^{-16}$ and $M_r/M_{\rm NS}=10^{-19}$, latter of which is taken in our previous calculation (e.g., \cite{2021ApJ...923...64D})}. Note that this condition tends to be relaxed during the NinjaSat phase because the burst calculations become highly unstable.

\section{Results}
\label{sec:res}

We calculate XRB models with the \texttt{HERES} code for the different sets of $(X/Y, Z_{\rm CNO}, \dot{M}_{-9})$, as described in the previous section. We compare the calculations of light curves in two different phases: (i) the early clocked burster phase (in Section~\ref{sec:res_clockd}) observed by INTEGRAL, NinjaSat, and NICER \citep{2024ApJ...968L...7N} and (ii) the decline phase of persistent luminosity (in Section~\ref{sec:res_decline}) observed by NinjaSat \citep{2025PASJL}.

As we cannot accurately estimate the bolometric luminosity from observed count rates, we conduct comparisons by shifting and scaling the observational data to align with the peak luminosity in our models. In the case of \ac{srga}, this treatment is justified according to \citet{2024A&A...690A.353M}; time evolution of count rates and bolometric flux, which are presented in their figures~3 and 4, respectively, show similar behavior, implying weak energy dependence on the energy conversion factor (see figures 2 and 3 in \cite{2024ApJ...968L...7N}).

Burst light curves can be generally characterized by several parameters (see \cite{2021ASSL..461..209G} for details). For the cases of solar composition and He-enhanced, we list some of the characteristic quantities in Appendix~\ref{app}.

\subsection{The clocked burst phase (early phase)}
\label{sec:res_clockd}

\begin{figure*}[t]
    \centering
    \begin{minipage}{0.45\linewidth}
        \includegraphics[width=0.9\linewidth]{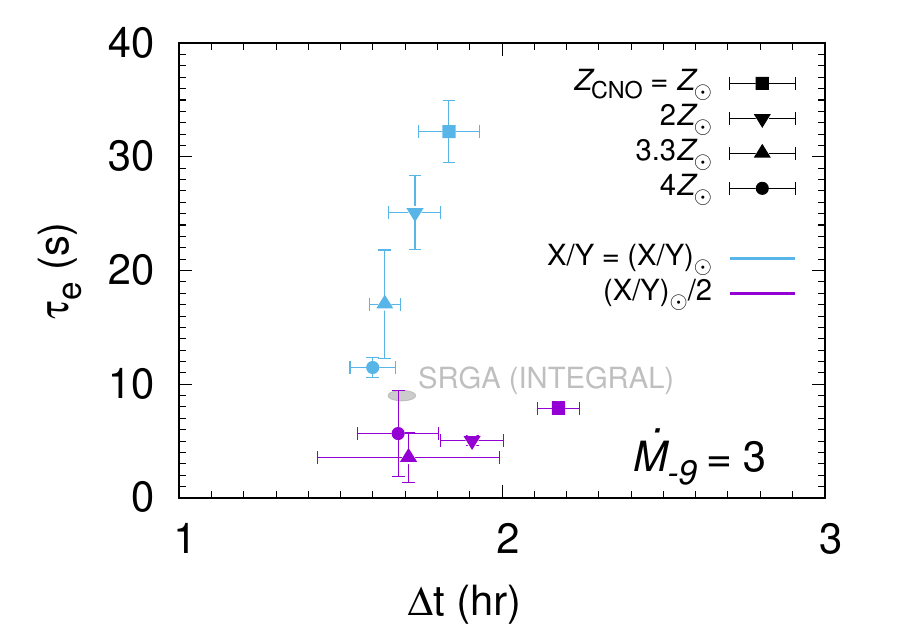}
    \end{minipage}
    \begin{minipage}{0.45\linewidth}
        \includegraphics[width=0.9\linewidth]{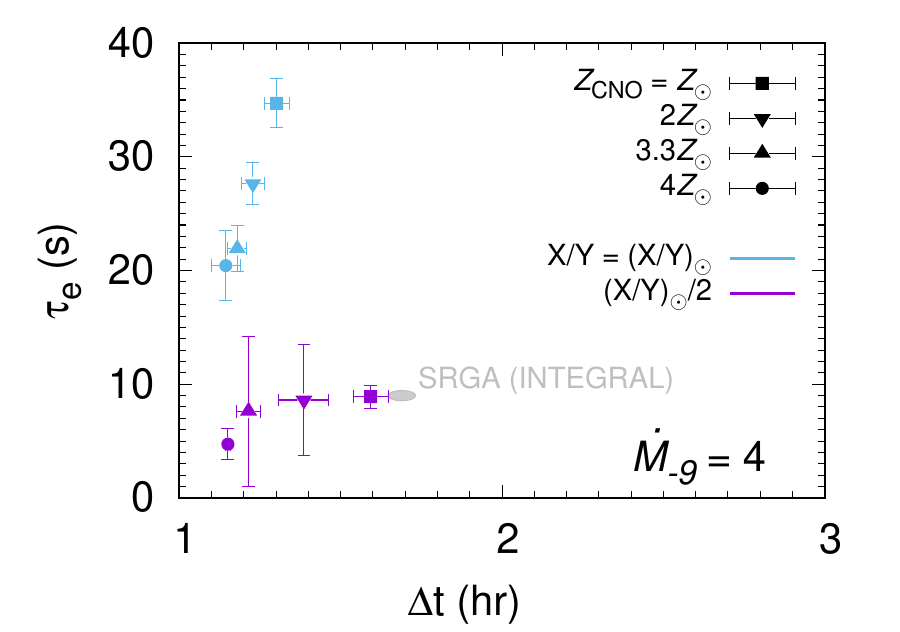}
    \end{minipage}  
    \caption{Calculated $\Delta t$ and $\tau_e$ with several $X/Y$ and $Z_{\rm CNO}$ values for different mass accretion rates, $\dot{M}_{-9} = 3$ (left), and $\dot{M}_{-9} = 4$ (right). The model errors are represented by the $1 \sigma$ error. The gray elliptical area with ``SRGA (INTEGRAL)'' indicates the reported value by \cite{2024ATel16485....1S}, assuming relative errors, 5\% for $\Delta t$ and 10\% for $\tau_e$. 
    }
    \label{fig:INTEGRAL}
\end{figure*}

\begin{figure*}[t]
    \centering
    \begin{minipage}{0.45\linewidth}
        \includegraphics[width=0.9\linewidth]{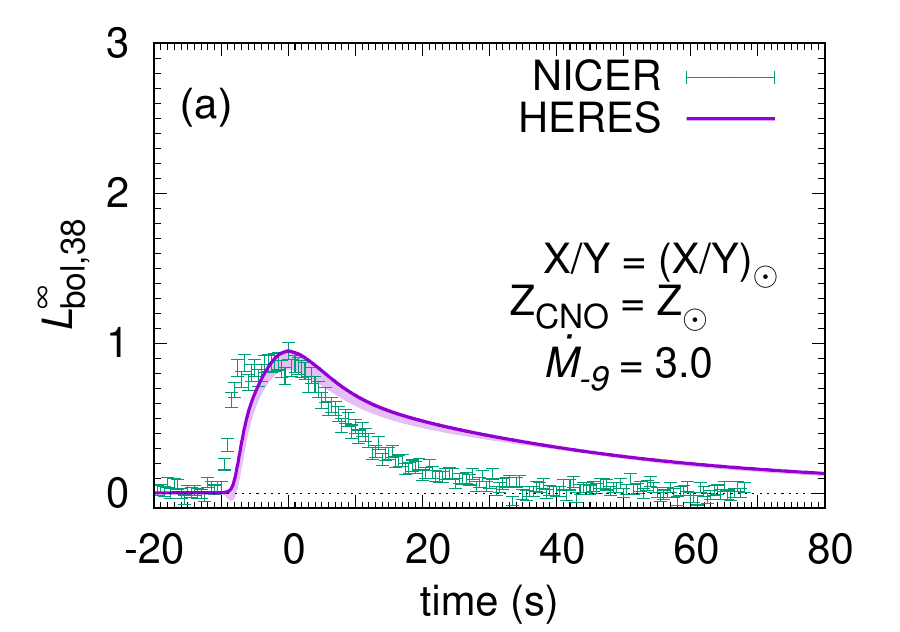}
    \end{minipage}
    \begin{minipage}{0.45\linewidth}
        \includegraphics[width=0.9\linewidth]{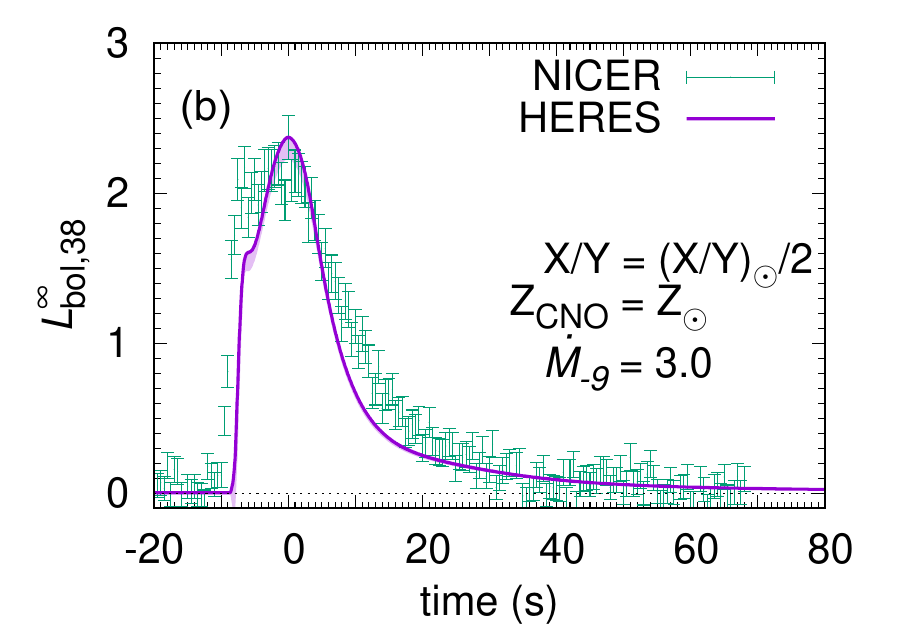}
    \end{minipage}
    \begin{minipage}{0.45\linewidth}
        \includegraphics[width=0.9\linewidth]{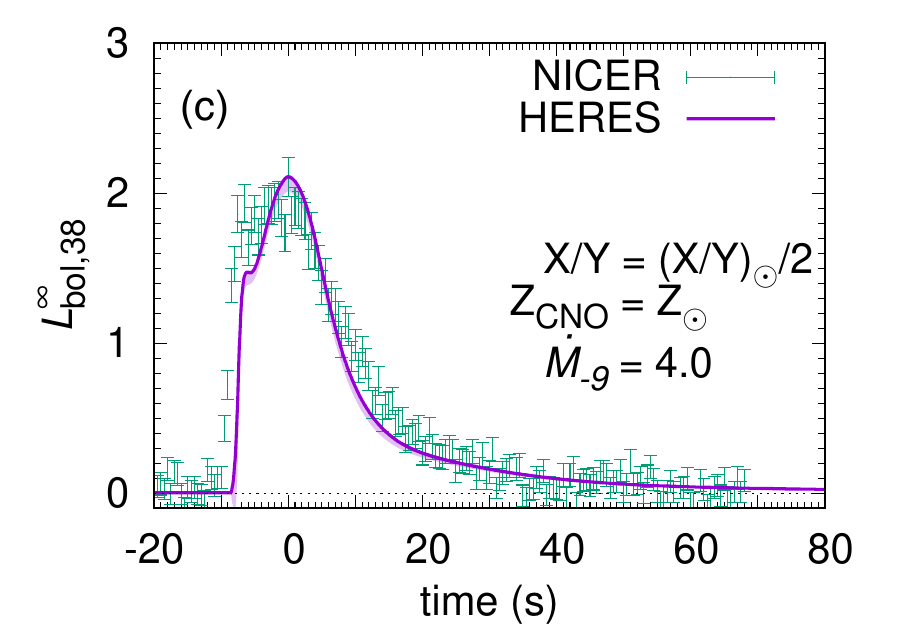}
    \end{minipage}
    \begin{minipage}{0.45\linewidth}        \includegraphics[width=0.9\linewidth]{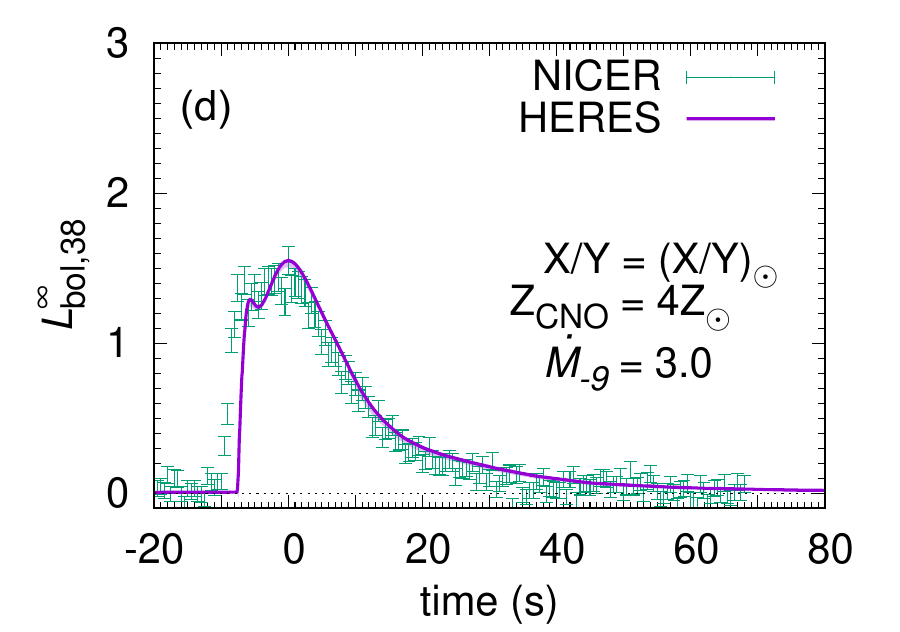}
    \end{minipage}
    \caption{Simulated light curves with different $X/Y$, $Z$, and $\dot{M}_{-9}$ compared with NICER observations \citep{2024ApJ...968L...7N}. (a) the solar composition, (b,c) He-enhanced case, and (d)
CNO-enhanced case.
     }
    \label{fig:nicer}
\end{figure*}

The calculated recurrence time $\Delta t$ and burst decay time $\tau_e$ are compared with that of \ac{srga} in figure~\ref{fig:INTEGRAL}, where we adopt the observed values reported by \citet{2024ATel16485....1S} as reference. In general, a lower $X/Y$ ratio ($0.5(X/Y)_\odot$, He enhanced) and/or a higher $Z_{\rm CNO}$ decrease $\tau_e$. For $\dot{M}_{-9} = 3$, most $0.5(X/Y)_\odot$ models, except those with $Z_{\rm CNO} = Z_\odot$, closely reproduce the observations. In particular, the metal-rich models with $Z_{\rm CNO} = 4~Z_\odot$ are favored compared to other compositions. In contrast, for higher mass accretion $\dot{M}_{-9} = 4$, all cases show a shorter $\Delta t$, with $X/Y = 0.5(X/Y)_\odot$ and $Z_{\rm CNO} = Z_\odot$ providing the closest match to the observations. In summary, the solar composition cases (i.e., $(X/Y,Z_{\rm CNO})=((X/Y)_\odot,Z_\odot)$ across all mass accretion rates appear unlikely to reproduce the observed values.

The comparison of light curve profiles also indicates a non-solar composition, as shown in figure~\ref{fig:nicer}: calculated light curves are plotted along with the NICER observation. In the solar composition case (figure~\ref{fig:nicer}a), values remain elevated from the peak through the tail due to nuclear heating from burning residual hydrogen \citep{1981ApJS...45..389W, 1984PASJ...36..199H}, as observed in bursts of the clocked burster GS 1826$-$24 \citep{2007ApJ...671L.141H}. This suggests that \ac{srga} may have a significantly lower remaining hydrogen abundance ($X$) after the rp-process, compared to previous bursters, probably due to an initially lower $X/Y$ ratio. In fact, the lower $X/Y$ models (figure~\ref{fig:nicer}b,c) closely match the observed light curve features after the peak.

The CNO-enhanced case (figure~\ref{fig:nicer}d) appears to reproduce the double-peak structure more effectively than the low $X/Y$ model. This double-peak structure may result from abundant hot CNO cycle seeds, as shown in \cite{2016ApJ...819...46L, 2024MNRAS.tmp..738S}. As the time scale for the luminosity increase shortens, the synthesis of the initial waiting point, ${}^{56}{\rm Ni}$, occurs earlier. Consequently, the rapid luminosity increase is moderated despite the increase in temperature in the ignition layer. Eventually, the luminosity reaches its peak, which is affected by the temporary stagnation of the rp-process before transitioning into the tail phase.

\subsection{The decline phase (later phase)}
\label{sec:res_decline}

\begin{figure*}[t]
    \centering
    \begin{minipage}{0.45\linewidth}
        \includegraphics[width=0.9\linewidth]{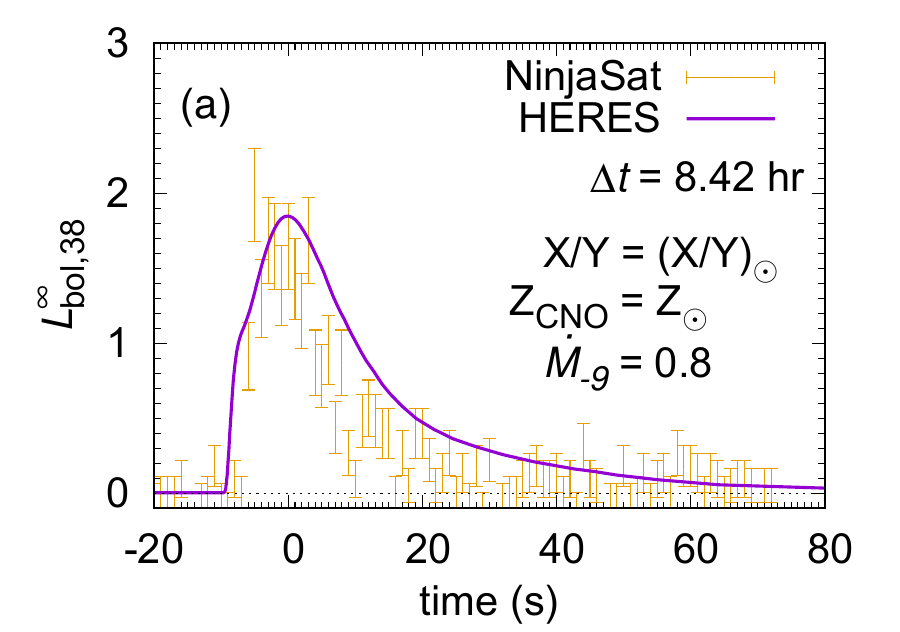}
    \end{minipage}
     \begin{minipage}{0.45\linewidth}
        \includegraphics[width=0.9\linewidth]{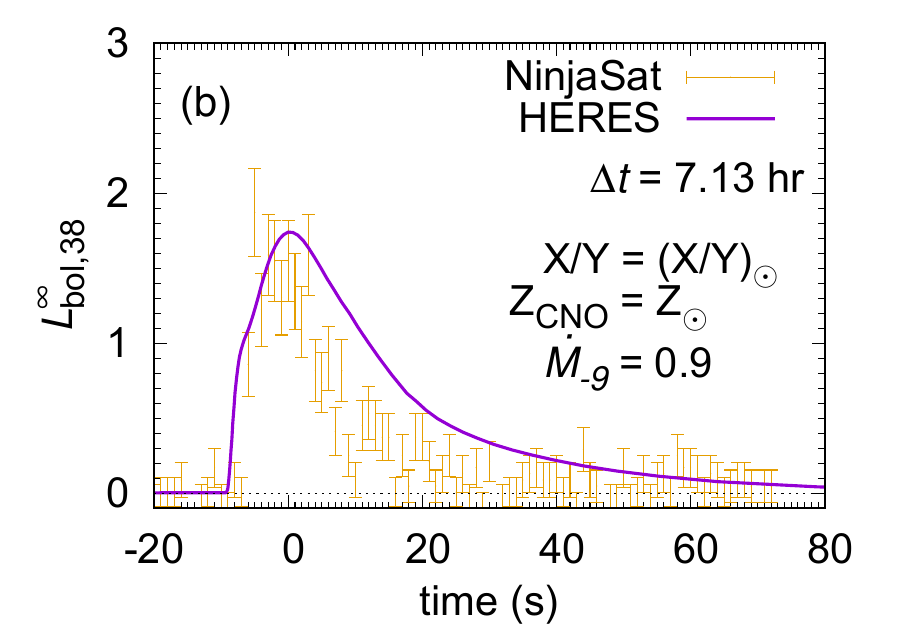}
    \end{minipage}
    \begin{minipage}{0.45\linewidth}
        \includegraphics[width=0.9\linewidth]{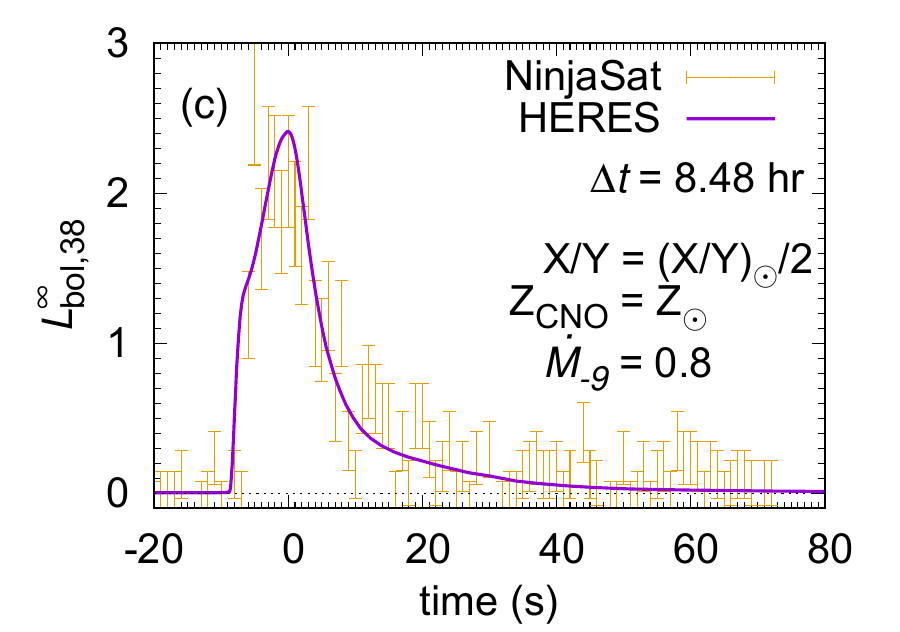}
    \end{minipage}
    \begin{minipage}{0.45\linewidth}
        \includegraphics[width=0.9\linewidth]{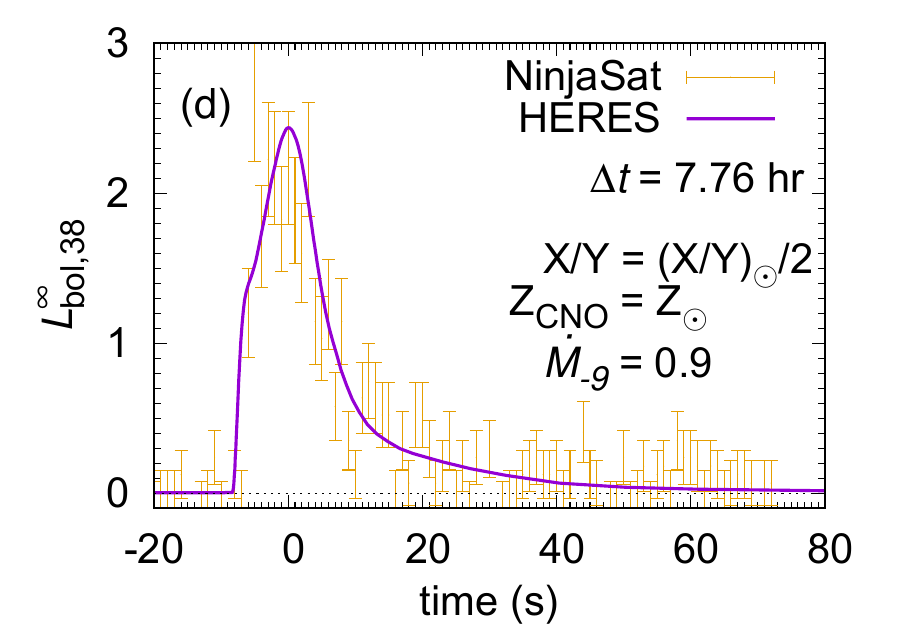}
    \end{minipage}
    \caption{Simulated light curves overplotted on observations by NinjaSat: (a,b) solar composition, (c,d) He-enhanced case.
    }
    \label{fig:ninjasat}    
\end{figure*}

For the later phase of \ac{srga}, we adopt the last two burst profiles of the calculation, where the accretion rate is lower by a factor of $\sim 4$ than other X-ray observations, as seen in figure~\ref{fig:obs}. 

Figure~\ref{fig:ninjasat} shows the results of simulated light curves along with observations by NinjaSat. The solar composition model (figure~\ref{fig:ninjasat}a and \ref{fig:ninjasat}b) has a longer burst duration than the observation. On the other hand, the He-enhanced model $X/Y=(X/Y)_\odot/2$ (figures~\ref{fig:ninjasat}c and \ref{fig:ninjasat}d), which has a shorter duration, shows better agreement. In particular, if we take $\dot{M}_{-9}=0.9$ (figure~\ref{fig:ninjasat}d), the corresponding recurrence time is $\Delta t = 7.76~{\rm h}$, close to the NinjaSat observation of $\Delta t = 7.909~{\rm h}$. Note that this He-enhanced model agrees with INTEGRAL and NICER with $\dot{M}_{-9,{\rm clocked}}=4$, which satisfies the constraint from persistent flux observations, equation~(\ref{eq:mdot}), in case of $\dot{M}_{-9,{\rm NinjaSat}}=4$. Thus, He-enhanced cases with solar metalicity, i.e., $X/Y\simeq(X/Y)_\odot/2$ and $Z_{\rm CNO}\simeq Z_\odot$, are preferred by \ac{srga} observations.

Regarding the metal-rich case with $Z_{\rm CNO} = 4~Z_\odot$, we conducted XRB calculations with $\dot{M}_{-9} = 0.8$ and $0.9$, both of which failed after the first XRB reached peak luminosity due to the hydrostatic limitations of \texttt{HERES}. This failure may imply the breaking of the hydrostatic equilibrium condition. Additionally, if we approximate the time since the start of the simulation to the first burst (when the simulation fails) as a lower limit to the recurrence time (see also \cite{2011PThPh.126.1177M} for a pure He-burning case), our \texttt{HERES} models predict $\Delta t = 25$ h and 12 h for $\dot{M}_{-9} = 0.8$ and $0.9$, respectively—both substantially longer than the observed $\Delta t$. Physically, this results from the extended duration of the hot CNO cycle, which efficiently consumes hydrogen that was supposed to be a fuel, sustained by a high abundance of CNO seeds. The low $\dot{M}_{-9}$ model, representative of a cooler environment, struggles to transition into a hot bi-CNO cycle through $\alpha$ capture on ${}^{14}{\rm O}$ and ${}^{15}{\rm O}$. Consequently, the hot CNO cycle could persist for an extended duration, a scenario considered in \citet{2017JPSJ...86l3901L} in relation to the extra heating source of MAXI J0556$-$332.

\begin{figure}[t]
    \centering\vspace*{-1cm}
         \includegraphics[width=0.8\linewidth]{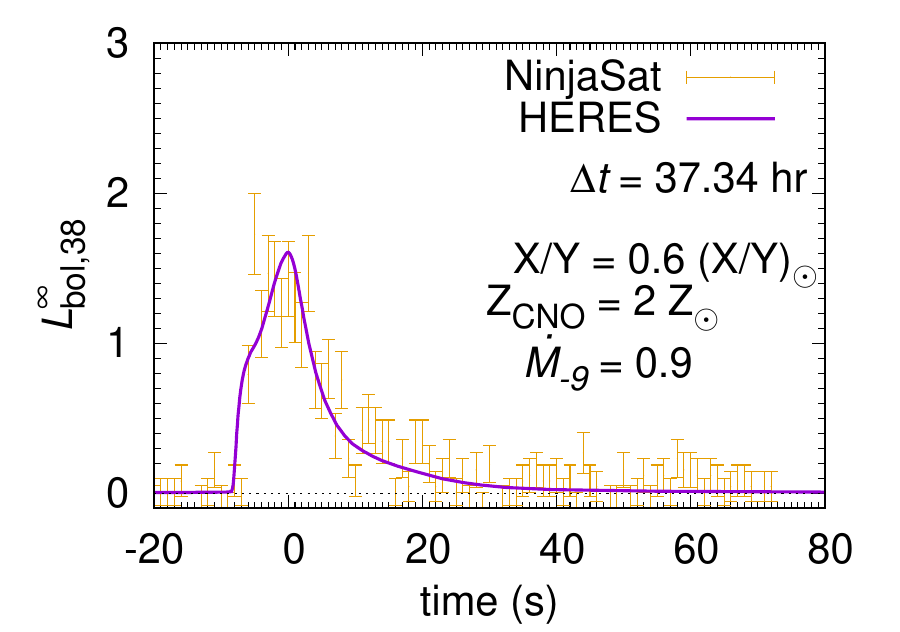}
    \caption{The marginally HeCNO-enhanced case with $X/Y=0.6~(X/Y)_\odot$, $Z_{\rm CNO}=2~Z_\odot$, and $\dot{M}_{-9}=0.9$.
  }
    \label{fig:HeCNO}
\end{figure}

We also consider cases where both He and CNO elements are enhanced. For example, we perform XRB calculations with $X/Y = 0.6~(X/Y)_\odot = 1.75$ and $Z_{\rm CNO} = 2Z_\odot$, representing a marginally HeCNO-enhanced scenario compared to our purely He- or CNO-enhanced models. During the clocked burster phase, this model produces results similar to the He-enhanced case, aligning well with INTEGRAL and NICER observations at $\dot{M}_{-9}=3.5$. In the decline phase, we set $\dot{M}_{-9} \lesssim 0.9$ based on equation(\ref{eq:mdot}), yielding a light curve that closely matches NinjaSat observations, similar to the He-enhanced scenario (figure~\ref{fig:HeCNO}). However, the recurrence time is too long ($\Delta t\simeq37~{\rm h}$) to match the observed 7.9 h due to a temperature low enough to delay the triple-$\alpha$ reaction, prolonging the hydrogen-triggered hot CNO cycle\footnote{At high $\dot{M}_{-9}$, the triple-$\alpha$ reaction initiates due to elevated temperature, without delay from the hydrogen-triggered hot CNO cycle. Testing the same composition at $\dot{M}_{-9}=1$ (unlikely for SRGA J1444 by equation~(\ref{eq:mdot})) yields $\Delta t=6.16~{\rm h}$, suggesting a critical $\dot{M}_{-9}$ between 0.9 and 1 that affects whether the hot CNO cycle quickly stabilizes in equilibrium.}. In short, the HeCNO-enhanced case, like the CNO-enhanced one, seems unfavorable because hydrogen accumulation needed for the XRB is difficult in low mass accretion scenarios for accreting NSs.

In summary, while various combinations of He and metal enhancement can replicate both the light curve profile and recurrence time during the clocked burster phase, increasing metallicity beyond the solar level tends to extend the recurrence time in the decline phase well beyond observed values. Thus, our current simulations by {\tt HERES} suggest that a helium-enhanced scenario is the most favorable for explaining the full range of observations by INTEGRAL, NICER, and NinjaSat.

\section{Summary and discussion}
\label{sec:summary}

In the present study, we calculated XRB models with the \texttt{HERES} code to investigate the behavior of the recently observed clocked burster \ac{srga}. By calculating with a wide range of model parameters, we have identified appropriate XRB model configurations that align with the recent multiple X-ray observations, such as the decay time (e-folding time) and recurrence time observed by INTEGRAL and the light curve profiles observed by NICER and NinjaSat. In particular, the short decay time and the plateau structure can be reproduced with He-enhanced models with $X/Y=1.5$, where rp-process heating is weaker. In this case, the corresponding accretion rate is obtained as $\dot{M}_{-9} \approx 3$--$4$ for the clocked phase, while $\dot{M}_{-9} \approx 0.8$--$0.9$ for the decline phase where the last two bursts were observed by NinjaSat.

The discovery of helium-enhanced XRBs provides valuable constraints on the preceding evolutionary history of the binary system. Specifically, it allows us to distinguish between channels that have descended from an intermediate mass X-ray binary as opposed to systems that had low-mass companions from birth \citep{2002ApJ...565.1107P}. In the former scenario, the outer layers of an initially intermediate mass ($\gtrsim1.5~\msun$) star possessing convective cores are stripped to expose deeper layers that used to be part of the convective core. Such layers will have elevated He abundances as well as CNO-processed metal abundances. In contrast, the latter scenario would predict the donor's surface composition to be similar to its primordial composition \citep{1996ApJ...458..301K}. It is the donor's surface material that accretes onto the NS, so it is clear that the two scenarios predict distinct accretion compositions. For the case of \ac{srga}, we constrain the hydrogen to helium ratio to be $X/Y\sim1.5$, corresponding to $X \sim 0.6$ and $Y \sim 0.4$. This is sufficiently helium-enriched compared to the solar composition ($Y\sim0.23$--$0.28$), supporting the intermediate mass X-ray binary descendant scenario. 

Together with other observational constraints, we can further constrain the companion star's birth mass within the intermediate mass descendant scenario. Since \ac{srga} is an X-ray pulsar, the mass function (relation between NS mass, its companion mass, and inclination angle) is known through pulsar timing models \citep{2024ApJ...968L...7N}. The inclination angle has been estimated to be $i=74.1^{+5.8}_{-6.3}~{}^{\circ}$ by recent IXPE's polarization observation~\citep{2024arXiv240800608P}. Combining these mass function and inclination angle constraints and assuming an NS mass of $M_{\rm NS}\gtrsim1.2 M_{\odot}$, we expect the current donor mass to be $M_\mathrm{don}\sim0.3$--$0.4~\msun$. Given that our XRB modeling indicates a modest but non-negligible deviation of the composition of the accreted matter from the solar value, we can infer that the current donor mass is just slightly smaller than the mass of the convective core at the zero-age main sequence ($M_\mathrm{conv,0}$), below which the composition becomes increasingly He-enhanced. Based on analytic models of convective core evolution \citep{2024arXiv240900460S}, we expect the mass difference to be within $1<M_\mathrm{don}/M_\mathrm{conv,0}\lesssim1.1$. Therefore, the initial convective core mass of the donor should have been slightly larger than the current donor mass $M_\mathrm{conv,0}\sim0.3$--$0.5~\msun$, roughly corresponding to an initial stellar mass of $M_\mathrm{don,0}\sim2$--$2.5~\msun$. These constraints can be further improved with more observations and detailed modeling, e.g., by comparing the radii of such partially stripped stars in detailed models and the observationally inferred Roche lobe size of the donor.

The high-$Z_{\rm CNO}$ scenario in our results in a longer recurrence time ($\Delta t = 25~{\rm h}$ for $\dot{M}_{-9} = 0.8$ and $\Delta t = 12~{\rm hr}$ for $\dot{M}_{-9} = 0.9$) compared to the NinjaSat observation, which shows $\Delta t \simeq 7.9~{\rm hr}$, suggesting an inconsistency. However, the recurrence time is significantly affected by the efficiency of the hot CNO cycle due to the presence of numerous CNO seeds, which is particularly sensitive to the \textit{bottleneck} reactions ${}^{14}{\rm O}(\alpha,p){}^{17}{\rm F}$ and ${}^{15}{\rm O}(\alpha,\gamma){}^{19}{\rm Ne}$ (e.g., \cite{2007ApJ...665..637F,2019ApJ...872...84M}). \cite{2018ApJ...860..147M} showed that, in the context of GS 1826$-$24, $\Delta t$ becomes around twice as short if the ${}^{15}{\rm O}(\alpha,\gamma){}^{19}{\rm Ne}$ is reduced by an order of magnitude. Thus, if the recurrence time $\Delta t$ can be shortened by reaction rate uncertainties (and possibly other model parameters), the high-$Z_{\rm CNO}$ scenario may not be entirely ruled out. Should the true reaction rates be higher than those used in the present study, which is plausible due to the absence of strict upper limits, the hot CNO cycle could proceed on a shorter timescale, thereby decreasing $\Delta t$. Furthermore, the simulated light curves during the clocked-burst phases match the observed features well, including the double-peak structure (see also discussion in \cite{2024MNRAS.tmp..738S}).

Astronomically, the question arises whether stars with such high metallicities exist in the Galactic outer disc region. Such high-$Z$ stars are commonly identified in the Galactic inner disc region \citep{2015ApJ...808..132H}, suggesting such stars may have migrated outward. \citet{2021ApJ...920L..32T} discusses the possibility of even further high metallicity stars $\sim 10~Z_\odot$ in the Galactic bulge transfer to the Solar system neighbor. Additionally, relaxing the stringent requirements for metal abundance would present a more frequent scenario. Therefore, further investigation into the high-$Z$ scenario for \ac{srga} still remains worthwhile from both nuclear physics and astrophysics perspectives.

In this study, we assume a canonical NS mass value of $1.4M_{\odot}$. However, we note that the exact value of the NS mass affects our constraints on the composition of $X/Y$ and $X_{\rm CNO}$ in accreting matter. As observational binary parameters of \ac{srga} improve, the NS mass estimate should become more precise. In fact, a significantly higher NS mass, potentially exceeding $2M_{\odot}$, is suggested by \citet{2025PASJL}--based on the flat-disk assumption \citep{1988ApJ...324..995F} and $\xi_b/\xi_p\simeq0.71$, derived by the IXPE observation \citep{2024arXiv240800608P}\footnote{This is due to a higher gravitational redshift, $z_g$, with $\xi_b/\xi_p \propto 1 + 1/z_g$ (see Section 6.4 in \cite{2018ApJ...860..147M}). However, the mass dependence of $\xi_b/\xi_p$ is likely more complex due to neutron star microphysics \citep{2021ApJ...923...64D}. Therefore, the correlation between the NS mass and the $\eta$ parameter in equation~(\ref{eq:eta}) \citep{2024ApJ...960...14D} is essential for accurately determining NS mass.}. A more systematic study of XRB light curves is desirable, especially in the case of massive NSs.

The analyses done in this Letter crucially hinge on the measurement of $\Delta t$ in the decline phase made by NinjaSat. Such persistent monitoring was only possible due to the flexibility in operations unique to small CubeSat satellites. These results demonstrate just a portion of the rich science that can be achieved from small-scale satellite observations in the future.

\begin{ack}
The authors thank S. Guichandut, H.L. Liu, S. Nagataki, Ph. Podsiadlowski, and T. Tsujimoto for their valuable discussions. This project was financially supported by JSPS KAKENHI (JP20H05648, JP21H01087, JP23KJ1964, JP23K19056, JP24H00008). N.N. received support from the RIKEN Intensive Research Project (FY2024-2025). Part of the computations in this work were performed using the computer facilities at CfCA in NAOJ and YITP at Kyoto University.
\end{ack}

\bibliographystyle{apj}
\bibliography{ref}

\appendix

\section{Burst characteristic quantities}
\label{app}
The burst light curves are characterized by several parameters, such as recurrence time ($\Delta t$), total burst energy in units of $10^{39}~{\rm erg}$ ($E_{b,39}$), peak luminosity in units of $10^{38}~{\rm erg~s^{-1}}$ ($L_{{\rm pk},38}$), and e-folding time ($\tau_e$). In table \ref{tab:table1} and \ref{tab:table2}, we list their values with $1\sigma$ errors in a number of burst profiles ($\hat{n}$) for the solar composition case ($X/Y=2.9$ and $Z_{\rm CNO}=0.015$) and the He-enhanced case ($X/Y=1.5$ and $Z_{\rm CNO}=0.015$), respectively, with different $\dot{M}_{-9}$. These values were utilized in \citet{2025PASJL} for comparison with observational data.

\begin{table}[h]
    \centering\vspace*{-0.25cm}
    \scalebox{0.75}{
    \begin{tabular}{c|ccccc}
$\dot{M}_{-9}$    & $\hat{n}$ & $\Delta t$ [h]& $E_{b,39}$ & $L_{{\rm pk},38}$ & $\tau_e$
[s]\\
\hline
0.8  & 1 &  8.42  & 4.12 & 2.37  & 6.3\\
 0.9  &1 &  7.13  &  4.70 & 1.74 & 2.2\\
 1    & 1 &  6.85  & 4.36 & 2.41 & 11.2\\
2  & 37 &    2.97$\pm$0.15    & 4.24$\pm$0.16 & 1.13$\pm$0.20 & 30.9$\pm$4.8 \\
2.5  & 61 &  2.30$\pm0.09$   & 4.20$\pm$0.18 & 1.04$\pm$0.14 & 31.8$\pm$2.9 \\
3    & 163 &  1.84$\pm$0.09  & 4.04$\pm$0.19 & 0.95$\pm$0.12& 32.6$\pm$2.8\\
 4  & 220 &   1.30$\pm$0.04     & 3.63$\pm$0.16 & 0.81$\pm$0.05& 34.5$\pm$2.2 \\
 5  & 221 &  0.97$\pm$0.05  & 3.11$\pm$0.20 & 0.71$\pm$0.06 & 34.4$\pm$2.2
 \end{tabular}
    }\caption{Several burst characteristics with solar compositions.}
    \label{tab:table1}
\end{table}

\begin{table}[h]
    \centering\vspace*{-1.0cm}
    \scalebox{0.75}{
    \begin{tabular}{c|ccccc}
$\dot{M}_{-9}$    & $\hat{n}$ & $\Delta t$ [h]& $E_{b,39}$ & $L_{{\rm pk},38}$ & $\tau_e$
[s]\\
\hline
 0.8  & 1 &  8.48  & 3.38 & 2.41  & 6.2\\
 0.9  & 1 &  7.76  &  3.63 & 2.44 & 7.4\\
 1    & 1 &  6.61  & 3.82 & 2.39 & 7.6\\
2  & 19 &    3.37$\pm$0.07    & 3.92$\pm$0.07 & 2.68$\pm$0.07 & 4.6$\pm$0.7\\
2.5  & 25 &  2.67$\pm$0.23  & 3.94$\pm$0.10 & 2.68$\pm$0.23 & 7.2$\pm$1.1\\
3    & 96 &  2.17$\pm$0.07  & 3.82$\pm$0.11 & 2.37$\pm$0.16 & 7.9$\pm$0.6 \\
 4  & 35 &   1.59$\pm$0.05     & 3.62$\pm$0.16 & 2.11$\pm$0.10 & 8.9$\pm$1.0 \\
 5  & 126 &   1.13$\pm$0.07  & 3.08$\pm$0.19 & 1.58$\pm$0.34 & 12.3$\pm$4.9
 \end{tabular}
    }\caption{Several averaged burst characteristics with He enhanced case.}
    \label{tab:table2}
\end{table}
\end{document}